\documentclass{ptapap}

\author{Ennio Poretti}[OAB]
\author{Nicolas Nardetto}[OCA]
\affil[OAB]{INAF-Osservatorio Astronomico di Brera, \\
Via E. Bianchi 46, 23807 Merate (LC), Italy}
\affil[OCA]
{Universit\'e C\^ote d'Azur, OCA, CNRS, Lagrange, Nice, France}

\title{Interferometric and spectroscopic observations
of the BRITE target $\delta$ Cep}

\begin{document}

\maketitle

\begin{abstract}

We used new HARPS-N spectra to revisitate the projection factor of
$\delta$~Cep and to directly measure the specific contribution of the
velocity gradient within the atmosphere.

\end{abstract}

\section{Introduction}
The classical Baade-Wesselink method allows us to measure  the diameters
of pulsating stars by using the radial velocity and colour curves over the pulsation period.
Instead of the colour curve, the interferometric version of the Baade-Wesselink approach
uses the angular variations of the stellar diameter, thus becoming a direct  method
to determine the distance of the pulsating star.
A few years ago we started a 
collaboration between the {\it Observatoire C\^ote d'Azur} (Nice, France) and
the {\it INAF-Osservatorio Astronomico di Brera} (Milano, Italy)  
aimed at improving the use of classical pulsators as stellar
candles. 
To pursue this goal we put together our expertises in stellar
interferometry and high-resolution spectroscopy \citep{guiglion, rhopup, vegachara}. 

\section{The physics behind the projection factor $p$}

\begin{figure}
\begin{minipage}{0.35\textwidth}
\centering
\includegraphics[width=\textwidth]{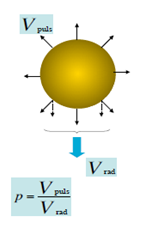}
\caption{Graphical representation of the projection factor $p$ between the
true pulsational velocity $V_{\rm puls}$ of the star and the radial velocity $V_{\rm rad}$
measured by the observer.}
\label{fig:p0}
\end{minipage}
\quad
\begin{minipage}{0.6\textwidth}
\centering
\includegraphics[width=\textwidth]{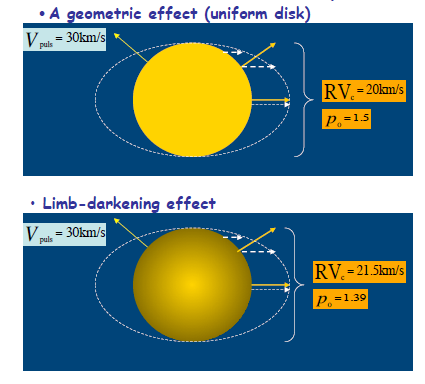}
\caption{The  limb-darkening effect (lower panel) decreases  the value
of the geometric factor (upper panel) due to the minor contribution of the
flux received from the borders of the stellar disc.}
\label{fig:p}
\end{minipage}
\end{figure}

When we measure the Doppler effect in the atmosphere of a pulsating star like
a Cepheid we actually measure the radial component of the pulsation velocity field,
i.e., the projection of the radial velocity along the line of sight
(Fig.~\ref{fig:p0}).
The recipe to go back to the true velocity pulsation $V_{\rm puls}$
from the observed $V_{\rm rad}$ contains three ingredients, reflecting three different physical
effects. They can be summarized in the decomposition of the projection factor 
$p=V_{\rm puls}/V_{\rm rad}$ into 
the geometrical factor $p_0$, the gradient $f_{\mathrm{grad}}$, and the correction factor
$f_{\mathrm{o-g}}$ \citep{nardetto2007}:

\begin{equation} \label{pfact}
 p= p_{\mathrm{o}}\,f_{\mathrm{grad}}\,f_{\mathrm{o-g}}.
\end{equation}

\begin{figure}
\centering
\includegraphics[width=0.75\textwidth]{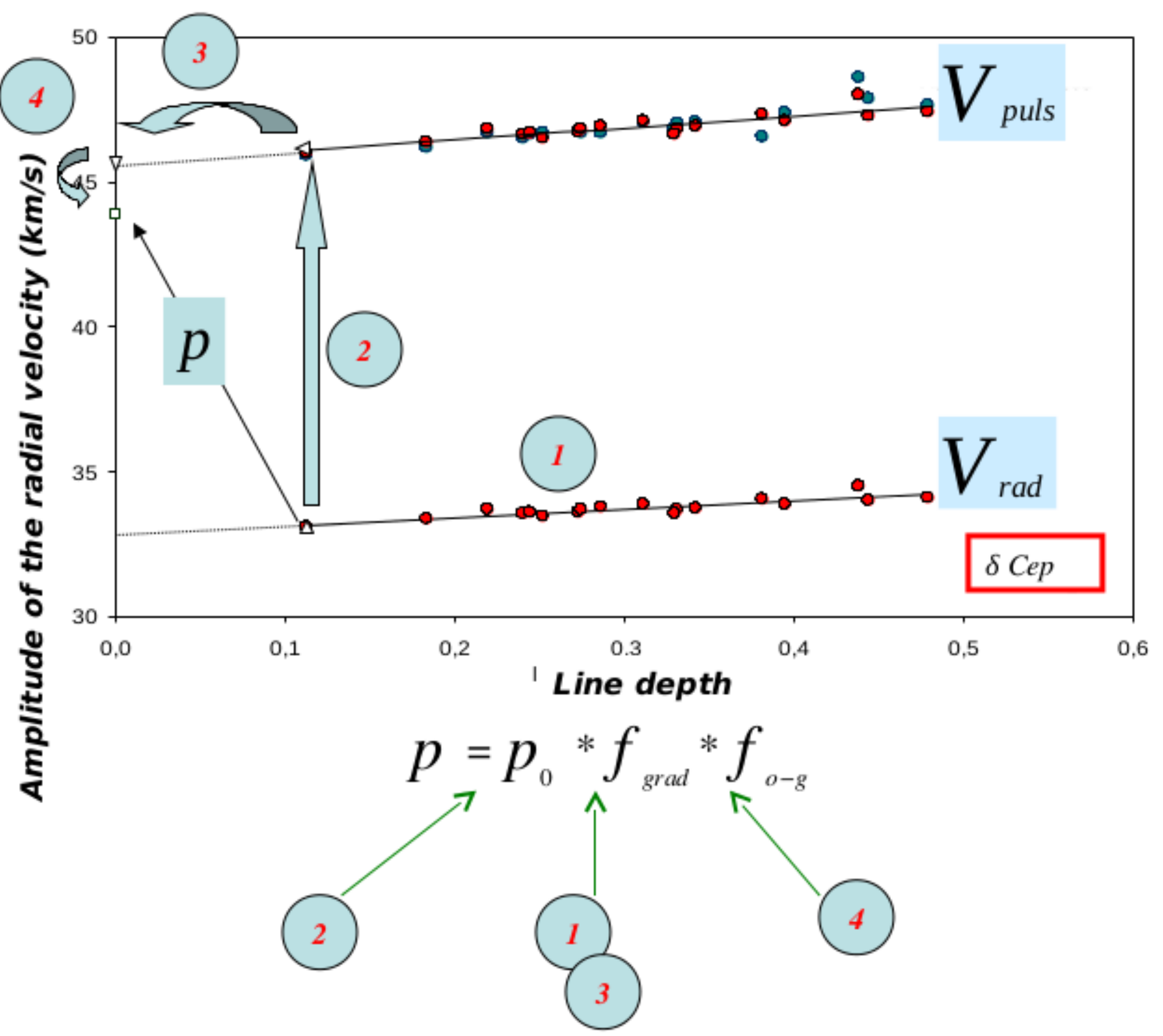}
\caption{Graphical representation of the decomposition of the projection
factor $p$: atmospheric gradient (line ``1"), geometric (shift ``2"), extrapolation
to the photosphere (``3"), and correction for the radial velocity of the gas
(shift ``4").}
\label{proj}
\end{figure}

The physical effects behind each factor are:

\begin{enumerate}
\item
we must  consider that the flux 
coming out from the borders is reduced by the limb darkening law, describing the
changes in the surface brigthness (Fig.~\ref{fig:p}, lower panel). 
Such a quantity acts a weight of the radial component
in function of the distance from the photocenter. 
The so-called geometrical factor
$p_{\mathrm{o}}$ has to be introduced (Fig.~\ref{proj}, shift indicated with ``2") to compensate this
effect; 
\item 
we measure the radial velocities of the lines in the spectra 
by means of their shifts in wavelength. We have to take into account that
the absorption lines are forming in different regions of the stellar atmosphere. 
In the extended atmospheres of Cepheids, different regions have different
radial velocity amplitudes and mean values. Therefore, such  values are line-dependent
and then prone to show a gradient  $f_{\mathrm{grad}}$ (Fig.~\ref{proj}, straight line
indicated with ``1");
\item 
the interferometric and photometric radii correspond to the photosphere of the star, 
hence we  have to extrapolate the spectroscopic one to it (hypothetical line of null depth; 
Fig.~\ref{proj}, indicated with ``3"). 
Moreover, spectroscopy is sensitive to gas ({\rm g}) movement while interferometry and photometry are
sensitive to the optical ({\rm o}) continuum. A last correction $f_{\mathrm{o-g}}$ is necessary 
to combine the different techniques (Fig. 3, shift indicated with ``4"). 

\end{enumerate}

It is quite evident that 
without knowing the projection factor $p$ we cannot use the 
Baade-Wesselink method to determine the distances of the sources and
of the environments in which they are embedded.

\section{The HARPS-N contribution}
The high-precision radial-velocity spectrograph HARPS-N \citep{co12} is installed at the 3.58-m {\it Telescopio Nazionale Galileo} (TNG), 
located at the Roque de los Muchachos Observatory (La Palma, Canary Islands, Spain).
One-hundred-three spectra of $\delta$ Cep were secured between March 27th and September 6th,  
2015 in the framework of the OPTICON proposal 2015B/015. Interferometric data of $\delta$ Cep were
previously obtained \citep{merand} with the Fiber Linked Unit for Optical Recombination 
\citep[FLUOR; ][]{fluor}
operating at the Center for High Angular Resolution Astronomy array
\citep[CHARA, Mount Wilson Observatory, USA; ][]{chara}. 
 By using the known trigonometric
distance of $\delta$~Cep \citep[$d$=272~pc; ][]{majaess} we could apply the {\it inverse} 
interferometric Baade-Wesselink method to the CHARA and HARPS-N data to derive an observed  
value of the projection factor $p$. We obtained $p_{\rm cc-g}$=1.239$\pm$0.031.

\begin{figure}
\includegraphics[width=\textwidth]{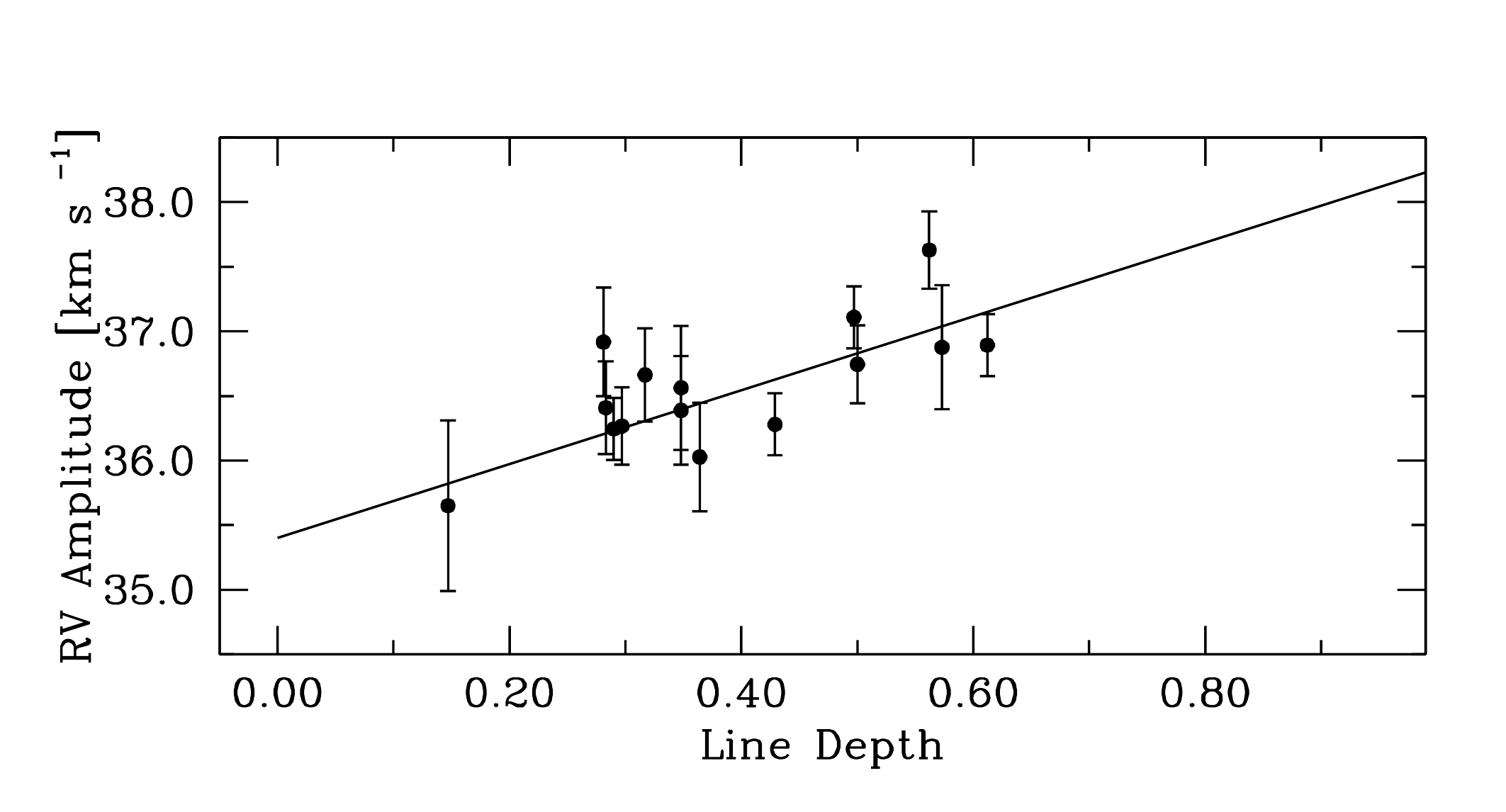}
\caption{HARPS-N variation of the amplitude of the radial velocity curve $\Delta RV_{\mathrm{c}}$ 
in function of the line depth $D$.}
\label{depth}
\end{figure}

In the context of the quantitative evaluation of the three physical effects
concurring to form the projection factor,
we used the HARPS-N spectra to derive the atmospheric gradient $f_{\rm grad}$ directly.
To do this we need a proxy for the height
of the line-forming region in the stellar atmosphere. Such proxy
is provided by the line depth $D$ taken at the minimum radius phase
\citep{nardetto2007}. The photosphere  sets the zero line depth. 
Figure~{\ref{depth}} shows the plot obtained from 15 unblended spectral lines
from 4683 to 6336~\AA\, \citep[twelve of Fe~{\sc i}, one each  of Ni~{\sc i}, 
Si~{\sc i} and Ti~{\sc ii}; two more lines of Fe~{\sc i} are not discussed here
for sake of semplicity; see ][ for further details]{harpsn}. We computed the amplitude of the radial velocity
curve $\Delta\,RV_c$ (the suffix $c$ stands for the centroid method, different
from  the
cross correlation $cc$ one used by the HARPS-N pipeline) 
for each line by means of a Fourier decomposition.
After this, we computed the  least-squares fitting line 

\begin{equation} \label{Eq_grad}
\Delta RV_{\mathrm{c}}= a_0 D + b_0 .
\end{equation}

The correcting factor $f_{\mathrm{grad}}$ due to the velocity gradient  can be
computed as the ratio between the amplitude extrapolated at the photosphere
($D$=0) and that of the given line \citep{nardetto2007}

\begin{equation} \label{Eq_grad2}
 f_{\mathrm{grad}}= \frac{b_0}{a_0D+ b_0}. 
\end{equation}

where $a_0$ and $b_0$ are the slope and zero-point of the linear fit. 
Therefore, for the first time we could determine 
$f_{\mathrm{grad}}$ from
the observations: the values are ranging from 0.964 (Fe~{\sc i} at 5367~\AA)
to 0.991 (Fe~{\sc i} at 4896~\AA),
with a typical error bar of 0.010. 

\section{Conclusions}
The first measurement of  $f_{\mathrm{grad}}$ was
not the only result obtained from HARPS-N spectra. 
By introducing an hydrodynamical model we could also determine the  semi-theoretical
values  $f_{\mathrm{o-g}}$=0.975$\pm$002 and $f_{\mathrm{o-g}}$=1.006$\pm$0.002 
by assuming radiative transfer in plane-parallel or sherically symmetric geometries,
respectively. 
The whole procedure is described in \citet{harpsn}. 

We are planning to observe other bright Cepheids both in interferometry
and high-resolution spectroscopy. 
An improving of the current performances
of the BRITE two-colours photometry could allow us to link the classical and
interferometric Baade-Wesselink method, thus
getting a closer look to the physics related to the projection factor and a
very important feedback on the distance scale \citep{spips}. This exercise would
be very useful to validate a new task merging interferometry, spectroscopy
and photometry to be included in the PLATO
complementary science, exploting at best the 
two onboard telescopes equipped with $B,V$ filters.

\acknowledgements{The observations leading to these results have received funding
from the European Commission's Seventh Framework Programme (FP7/2013-2016)
under grant agreement No. 312430 (OPTICON). 
EP and NN acknowledge financial support from the PRIN-INAF 2014 and the ANR-15-CE31-0012- 01, respectively.
Based in part on data collected by the BRITE-Constellation satellite mission, built, launched 
and operated thanks to support from the Austrian Aeronautics and Space Agency (FFG-ALR) 
and the University of Vienna, the Canadian Space Agency (CSA), and the Foundation for 
Polish Science \& Technology (FNiTP MNiSW) and National Science Centre (NCN).
}

\bibliographystyle{ptapap}
\bibliography{Poretti}

\end{document}